\documentclass[cits]{PoS}

\title{Taylor- and fugacity expansion for the effective center model of QCD at finite density}

\ShortTitle{Taylor- and fugacity expansion for the effective center model of QCD at finite density}

\author{\speaker{Eva Gr\"unwald}
\\Institut f\"ur Physik, Universit\"at Graz\\
E-mail: \email{eva$\_$gruenwald@gmx.at}}

\author{Ydalia Delgado Mercado\thanks{Supported by Strongnet, HP2 and TRR 55.}\\
\\Institut f\"ur Physik, Universit\"at Graz\\
E-mail: \email{ydalia.delgado-mercado@uni-graz.at}}

\author{Christof Gattringer
\\Institut f\"ur Physik, Universit\"at Graz\\
E-mail: \email{christof.gattringer@uni-graz.at}}

\abstract{Using the effective center model of QCD we test series expansions for finite chemical potential $\mu$. In particular we study two 
variants of Taylor expansion as well as the fugacity series. The effective center model has a dual representation where the sign problem is 
absent and reliable Monte Carlo simulations are possible at arbitrary $\mu$. We use the results from the dual simulation as reference data 
to assess the Taylor- and fugacity series approaches. We find that for most of 
parameter space fugacity expansion is the best (but also numerically most expensive) choice for reproducing the dual simulation results, while conventional Taylor 
expansion is reliable only for very small $\mu$. We also discuss the results of a modified Taylor expansion in $e^{\pm \mu} - 1$ which at the same numerical effort 
clearly outperforms the conventional Taylor series. } 
\FullConference{31st International Symposium on Lattice Field Theory LATTICE 2013\\
July 29 $ - $ August 3, 2013\\
Mainz, Germany}

\begin{document}

\section{Introduction}
Lattice QCD at non-zero chemical potential $\mu$ suffers from the complex action problem which makes finite density lattice QCD inaccessible to 
conventional Monte Carlo techniques. An alternative strategy that has been explored 
are expansions around the $\mu = 0$ theory using, e.g., Taylor- or fugacity series. In this contribution
we study fugacity expansion, Taylor series and a modified Taylor series in a QCD related model, referred to as the effective center model 
\cite{centermodel1,centermodel2}. The effective center model is an effective theory for the dynamics of the Polyakov loops 
\cite{centerbreaking} and contains a center symmetric interaction between two $\mathbb{Z}_3$-valued Polyakov spins on nearest neighbors
and the leading center symmetry breaking term from the fermion determinant, which also couples to the chemical potential. The effective center model can be
mapped exactly to a dual representation where the complex action problem is solved and reliable Monte Carlo simulations are possible for arbitrary $\mu$. We 
use the results from the dual representation as reference data to assess the reliability and convergence region of the three series we study.

The $\mathbb{Z}_3$ effective model may be derived from full QCD using a strong coupling 
approximation for the gluon action and a hopping expansion for the fermion 
determinant. Moreover, exploring the Svetitsky-Yaffe 
conjecture \cite{centerbreaking}, the degrees of freedom are reduced to elements of the center group  $\mathbb{Z}_3$. 
The action of the $\mathbb{Z}_3$ effective model reads
\begin{equation}
S_\mu \;  = \; - \sum_x \left[ \tau  \sum_{\nu = 1}^{3} \left[ P_x P_{x + \hat{\nu}}^*  + c.c. \right] + \eta P_x + \bar{\eta}P_x^* \right] ,
\label{Z3effectiveaction}
\end{equation}
where the $P_x$ are elements of $\mathbb{Z}_3 = \left\lbrace 1, e^{\pm 2 i \pi /3} \right\rbrace$. 
The first sum runs over all sites $x$ of a $N^3$ lattice with periodic boundary conditions and 
$\hat{\nu}$ denotes the unit vector in $\nu$-direction. The chemical potential $\mu$ enters through $\eta = \kappa e^\mu $, 
$\bar{\eta} = \kappa e^{ -\mu}$. The parameter $\tau$ is increasing with temperature, whereas $\kappa$ is increasing with decreasing QCD quark mass and is 
proportional to the number of flavors. The partition function is obtained as a sum over all configurations $\{ P \}$ of the variables, $Z = \sum_{\{ P \}} e^{-S_\mu}$.
In this study we focus on the expectation value $\langle P \rangle = V^{-1} \langle \sum_x P_x \rangle  = V^{-1} \partial \ln Z / \partial \eta $ 
of the Polyakov loop and the 
corresponding Polyakov loop susceptibility as our main observables. 

It is obvious that in the standard representation the action (\ref{Z3effectiveaction}) is complex for $\mu \neq 0$ and conventional Monte Carlo techniques
fail. The exact transformation to a flux representation solves the complex action problem for this model and allows for the 
application of Monte Carlo techniques at arbitrary 
$\mu$. In \cite{centermodel2} the phase diagram in the $\mu$-$\tau$ plane was mapped out and 
we use the results from that study as reference data for the analysis
of our series expansions. 

\begin{figure}
\begin{minipage}[b]{0.49\linewidth}
\centering
\includegraphics[height=53mm,clip]{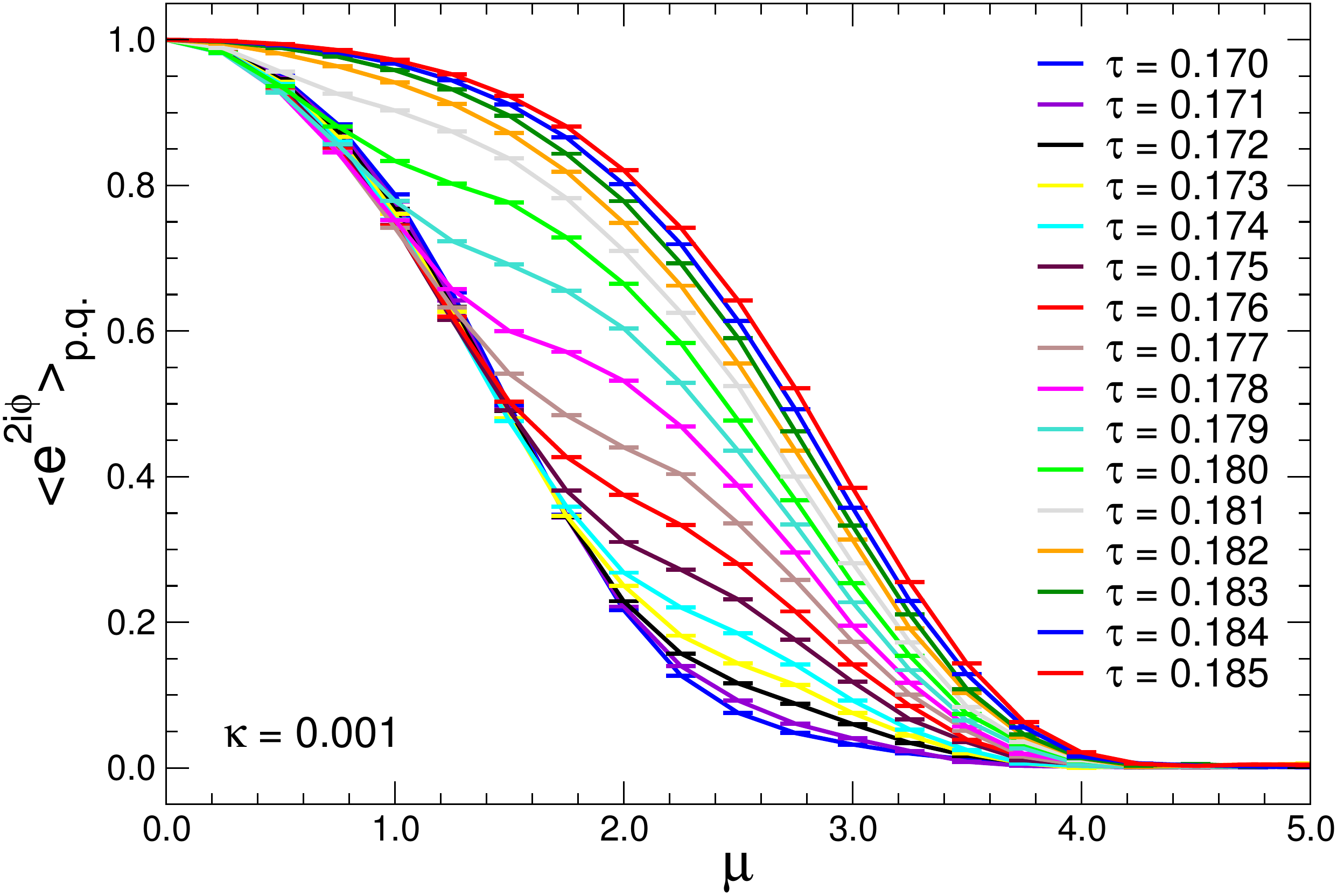}
\label{fig:figure22}
\vspace*{-5.2mm}
\end{minipage}
\hspace{4mm}
\begin{minipage}[b]{0.49\linewidth}
\centering
\includegraphics[height=53mm]{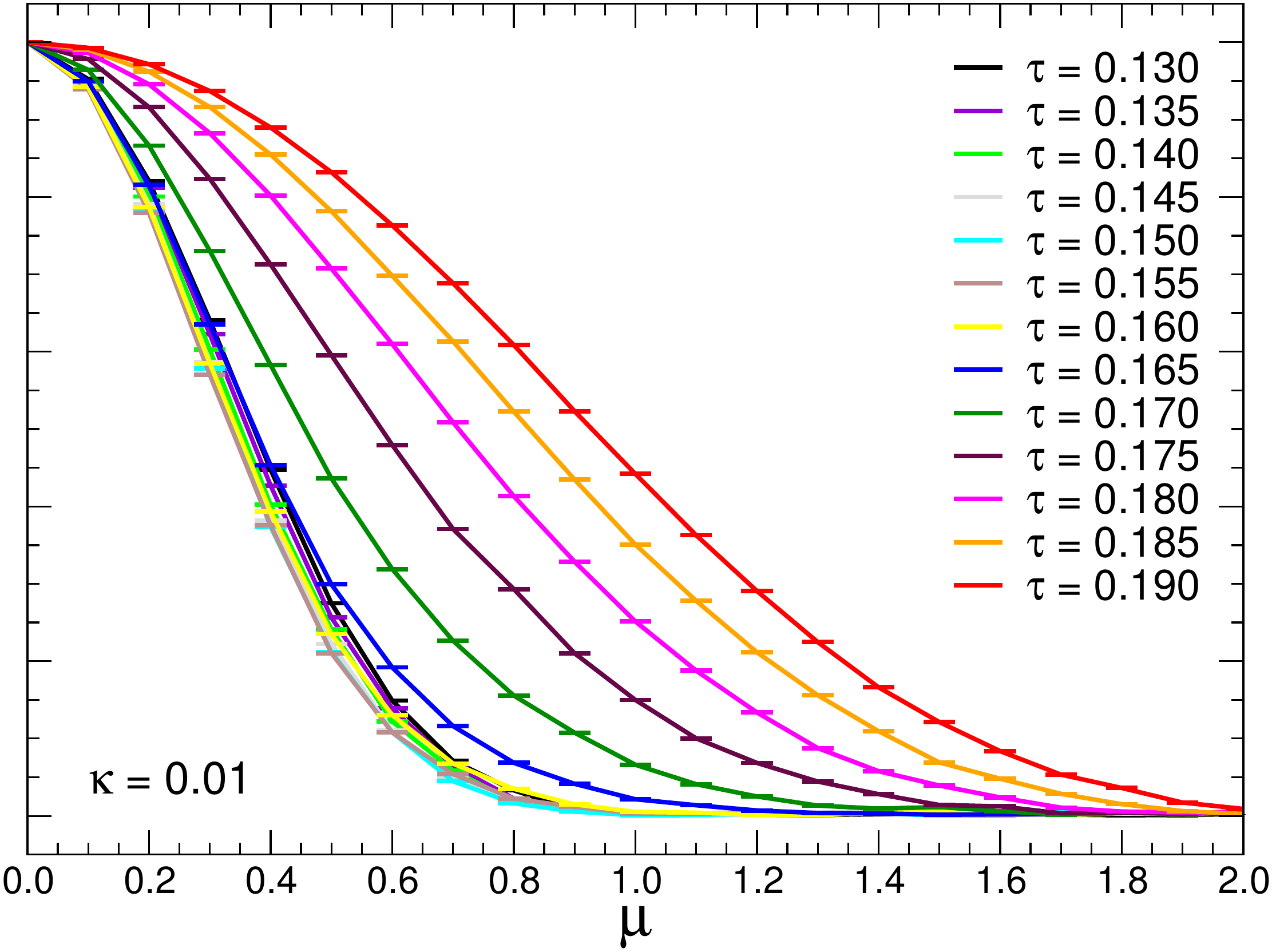}
\label{fig:figure23}
\end{minipage}
\caption{$\left\langle e^{i2 \phi} \right\rangle _ {p.q.}$ versus $\mu$ in the phase quenched theory for $\kappa = 0.001$ (lhs.\ plot) and 
$\kappa = 0.01$ (rhs.) on $16^3$ lattices for different values of $\tau$. The smaller $\tau$, the further left is the corresponding curve. Note the different 
scale on the horizontal axes of the two plots.}
\label{CAP}
\end{figure}

For later use in the discussion of our results for the various series expansions, in Fig.~\ref{CAP} we assess the severity of the complex action 
problem by studying the phase $e^{i\phi}$ of the Boltzmann factor in the phase quenched theory. 
We show plots of  $\left\langle e^{i2 \phi} \right\rangle _ {p.q.}$ 
as a function of $\mu$ for $\kappa = 0.001$ (lhs.\ plot) and $\kappa = 0.01$ (rhs.) on $16^3$ lattices for different values of $\tau$.

From Fig.~\ref{CAP} it can be seen, that for higher values of $\kappa$ (corresponding to smaller particle mass) the complex action problem is already
severe at smaller values of the chemical potential and that smaller temperatures are more affected. Since $\tau_{crit}$ shifts to smaller values of the
temperature $\tau$ when the chemical potential is increased, this fact complicates the exploration of the 
phase diagram when using series expansion methods.

\section{Fugacity expansion of the partition function}
We begin our study with the fugacity expansion 
\begin{equation}
Z \;  = \; \sum_{q \in \mathbb{Z}} e^{\mu q} \, Z_q \; ,
\label{equation:Z_fug}
\end{equation}
where the sum runs over all net particle numbers $q$. The canonical partition sums $Z_q$ are given by
\begin{equation}
Z_{q} \; = \; \sum_{\{P\}} e^{\tau  \sum_{x,\nu} \left[ P_x  P_{x + \nu}^* + c.c. \right]  } \, D_q \; ,
\end{equation}
where the $D_q$ are the analogues of the canonical determinants of QCD, i.e., the fermion determinant projected to a fixed quark number sector. 
As the canonical determinants in QCD \cite{qcdcandet}, the
$D_q$ can be computed as Fourier transforms with respect to imaginary chemical potential and are given by  ($\; H \equiv \sum_x P_x = R e^{i \theta }\,$)  
\begin{equation}
D_{q} \; = \;  \frac{1}{2\pi} \int_{-\pi}^\pi d\varphi \,   e^{-i\varphi \, q} \; \exp \left( \kappa e^{i\varphi} H + \kappa e^{-i\varphi} H^* \right)  \; = \; 
 e^{i \theta \, q}  I_q \left( 2 \kappa R \right) ,
\end{equation}
where $I_q$ denotes the modified Bessel functions. 

We begin our analysis of the properties of the fugacity expansion by inspecting the modulus of the coefficients $D_q$ as a function of $q$.
It is obvious that the $D_q$ must decrease with increasing $q$, 
such that the fugacity series (\ref{equation:Z_fug}) converges. In a practical implementation 
the fugacity series must be truncated to values $q$ of the particle number in some interval with a lower and an upper bound, i.e., $q_l \leq q \leq q_u$. The analysis 
of the size distribution of the $D_q$ is necessary for obtaining a reasonable estimate for $q_l$ and $q_u$. 

\begin{figure}[t]
\begin{minipage}[b]{0.49\linewidth}
\centering
\includegraphics[width=\textwidth]{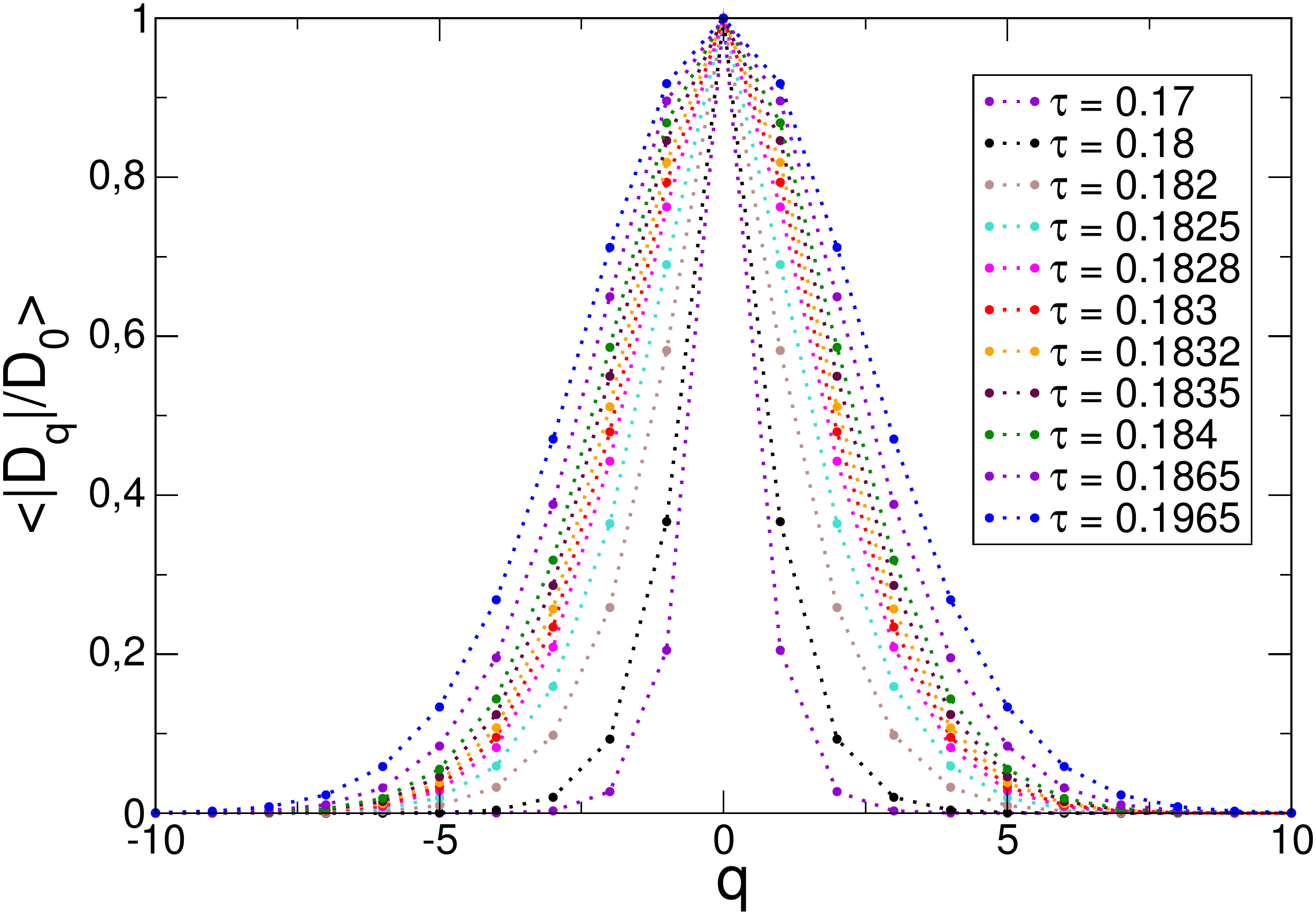}
\label{fig:figure20}
\vspace*{-0.5cm}
\end{minipage}
\hspace{0.1cm}
\begin{minipage}[b]{0.49\linewidth}
\centering
\includegraphics[width=\textwidth]{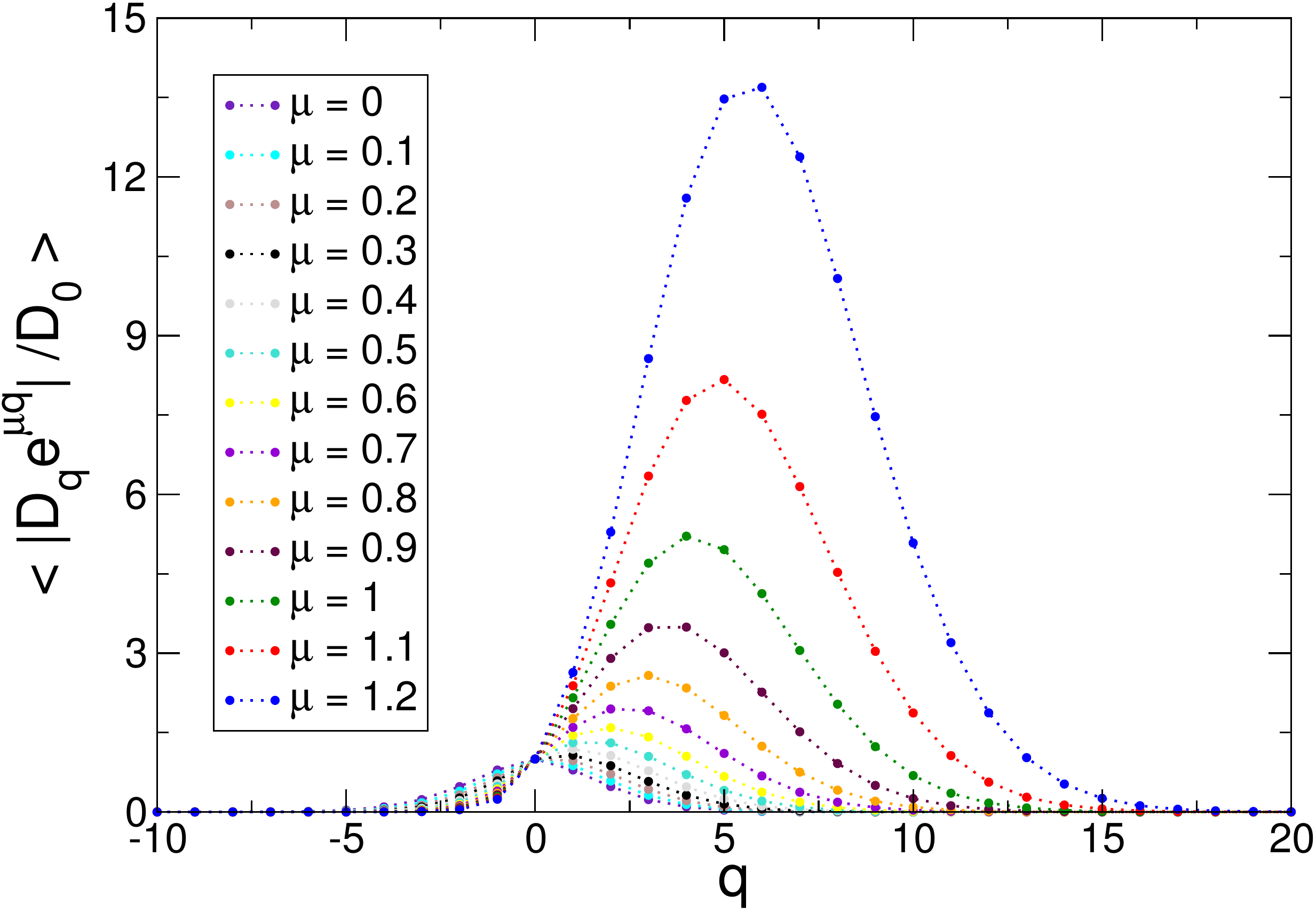}
\label{fig:figure21}
\vspace*{-0.5cm}
\end{minipage}
\caption{Distribution of the coefficients in the fugacity series. In the lhs.~plot we show  $\langle | D_q | / D_0 \rangle$ versus $q$ 
at $\kappa = 0.001, \mu = 0$ for  $16^3$ lattices for different values of temperature parameter $\tau$ (width of distribution increases with $\tau$). 
On the rhs.\ we show $\langle | e^{\mu q} D_q | / D_0 \rangle$ for $\kappa = 0.001, \tau = 0.183, 16^3$ for different values of $\mu$.}
\label{gauss_CAP}
\end{figure}

In the lhs.~plot of Fig.~\ref{gauss_CAP} we show the expectation value $\langle | D_q | / D_0 \rangle$ versus $q$ at $\kappa = 0.001$ ($\mu = 0$)  
for different values of $\tau$ on $16^3$ lattices. The distribution has a Gaussian-like shape, 
with the width of the distribution increasing with the temperature 
parameter $\tau$. This behavior is to be expected, 
since the width of the distribution is related to the particle number susceptibility which increases with $\tau$. 
The analysis shows that on the $16^3$ lattices for all values of $\tau$ we consider, the main contributions to the $\mu = 0$ 
fugacity series are taken into account 
for $q_l = -10$, $q_u = +10$. 
 
The chemical potential enters the fugacity series via the factor $e^{\mu q}$, shifting the $D_q$ that contribute to the fugacity expansion towards
larger values of $q$. This is evident from the rhs.\ plot in Fig.~\ref{gauss_CAP},  where we show $\langle | e^{\mu q} D_q | / D_0 \rangle$ versus $q$ for
$\kappa = 0.001, \tau = 0.183$ for different values of $\mu$. For the range of chemical potential values considered here a reasonable choice for
the truncated series would be $q_l = -5$, $q_u = +20$. 

Of course the analysis in Fig.~\ref{gauss_CAP} is only of a qualitative nature and is presented here to illustrate the effects of the interplay between 
the chemical potential and the size distribution of the  $| D_q |$. The optimal truncation parameters $q_l$ and $q_u$ were determined by 
systematically studying the relative error between the exact expression and the truncated series as a function of $q_l$ and $q_u$.

A detailed comparison of the fugacity expansion results for physical observables to our reference data from the dual representation 
will be presented in Section~5. 
However, we already remark now that for most of the parameter values studied here, the fugacity 
expansion results agree very well with the reference curves.

\section{Regular Taylor expansion (RTE) in the effective $\mathbb{Z}_3$ model}
The observables we consider here can be obtained as derivatives of the logarithm of the partition function. This logarithm now is Taylor-expanded in $\mu$ and 
derivatives at $\mu = 0$ are the coefficients of the resulting series (which of course is truncated in an actual application),
\begin{equation}
\ln Z  = \sum_{n = 0}^\infty  \frac{\mu^n}{n!}  \left(\! \frac{\partial}{\partial \mu}\! \right)^n \, \ln Z  \, \bigg|_{\mu = 0} \; .
\end{equation}

\begin{figure}[ht]
\begin{minipage}[b]{0.49\linewidth}
\centering
\includegraphics[height = 55mm]{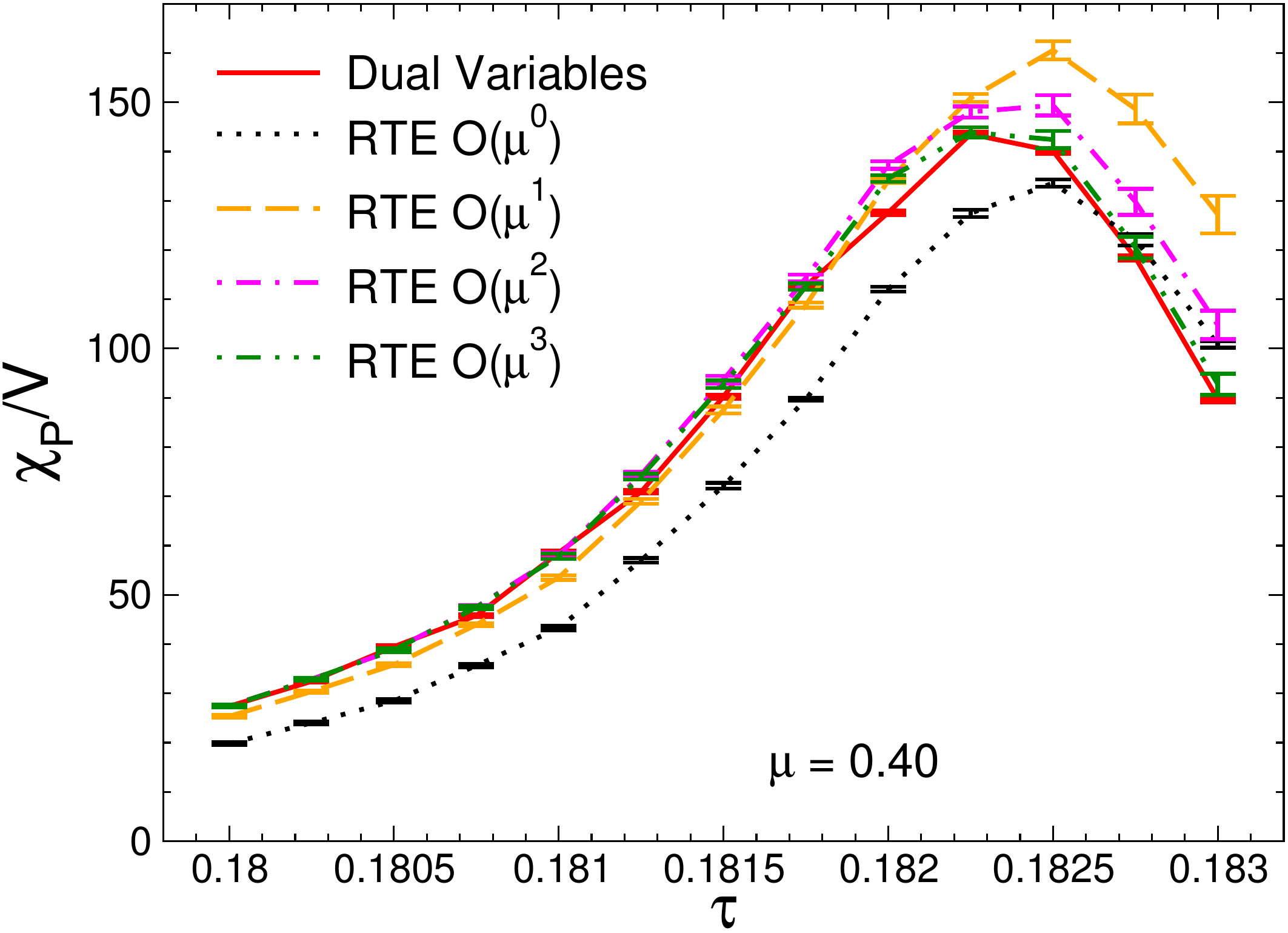}
\label{fig:figure20}
\vspace*{-0.0cm}
\end{minipage}
\hspace{0.3cm}
\begin{minipage}[b]{0.47\linewidth}
\centering
\includegraphics[height = 55mm]{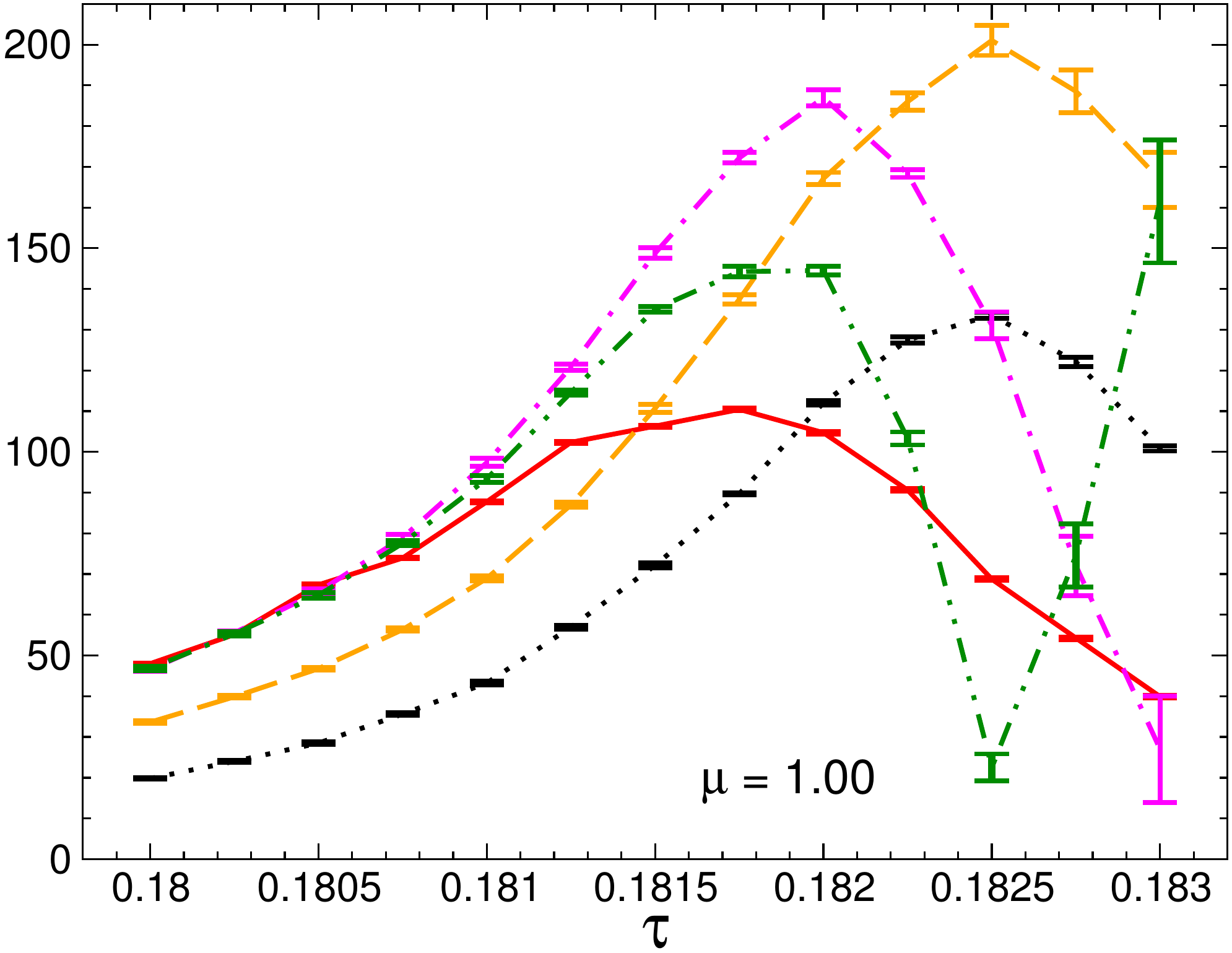}
\label{fig:figure21}
\end{minipage}
\caption{Polyakov loop susceptibility obtained from the regular Taylor expansion for $\kappa = 0.001$, $\mu = 0.4$ (lhs.\ plot) and $\mu = 1.0$ (rhs.)
as a function of the temperature. Results up to third order in $\mu$ are shown and are compared to flux representation 
results on a $16^3$ lattice.}
\label{fig:chiP_k0001_RTE_dual}
\end{figure}

In Fig.~\ref{fig:chiP_k0001_RTE_dual} we compare the Taylor expansion results up to third order in $\mu$ for the Polyakov loop susceptibility to
the results obtained in terms of dual variables. We use $16^3$ lattices at $\kappa = 0.001$, $\mu = 0.4$ (lhs.\ plot) 
and $\mu = 1.0$ (rhs.) for a statistics of $10^6$ 
configurations. 
The plot demonstrates that in general the Taylor series reproduces the dual results better for the smaller values of the temperature parameter $\tau$. 
At a chemical potential of $0.4$ we can achieve quite good agreement with the dual results for the whole $\tau$ range when we include all terms up to 
$\mathcal{O}(\mu^3)$. For $\mu = 1.0$, however, it is obvious that the Taylor series up to  $\mathcal{O}(\mu^3)$ 
fails to reproduce our reference data except 
for the smallest values of $\tau$.

\section{Improved Taylor expansion (ITE) in the effective $\mathbb{Z}_3$ spin model}
We now consider a second type of Taylor series, which we refer to as the ''improved Taylor expansion'', 
where the logarithm of the partition sum is expanded 
in the parameters $ \rho = \kappa \left(  e^{\mu} - 1  \right)$ and $\ \bar{\rho} = \kappa \left(  e^{- \mu} - 1  \right)$ 
(which in the limit $\mu \rightarrow 0$ corresponds to an expansion in $\mu$). 
A part of the motivation for this type of expansion is to capture some of the features of the fugacity expansion, which in the case of QCD would lead
to a finite Laurent series, whereas the regular Taylor expansion gives rise to an infinite series.
For the ITE the Boltzmann factor is organized as follows, 
\begin{equation}
e^{- S_{\mu}} \; = \; e^{-S_0} \; e^{\rho H + \bar{\rho}H^*},
\end{equation}
where  $S_\mu$ is the action as given in (\ref{Z3effectiveaction}) and $S_0$ its form when $\mu = 0$. 
When expanding the second factor on the rhs.\ one may
express the partition sum $Z(\mu)$ at non-zero $\mu$ in the following series,
\begin{equation}
Z(\mu) \; = \;Z(0) \, \left[ 1 + \left\langle \rho H + \bar{\rho} H \right\rangle_0 + 
				      \frac{1}{2}  \left\langle \left(  \rho H + \bar{\rho} H  \right)^2  \right\rangle_0  +
				    \frac{1}{6}  \left\langle \left(  \rho H + \bar{\rho} H  \right)^3 \right\rangle_0 + \mathcal{O} \left( \rho^4 \right)
				      \right] .
\end{equation}
The individual terms are expectation values $\langle .. \rangle_0$ evaluated in the $\mu = 0$ theory. They have a structure different from the terms
in the regular Taylor series but their evaluation in full QCD has the same numerical cost as the coefficients of the RTE.
The logarithm of the partition function for the evaluation of observables is obtained by further Taylor expansion in $\rho$ and $\bar{\rho}$ and 
observables by subsequent derivatives. 

Figure \ref{fig:chiP_k0001_ITE_dual} shows the Polyakov loop susceptibility at $\kappa = 0.001$ as a function of the temperature for two values of $\mu$
and again we compare the series for orders up to $\mathcal{O}(\rho^3)$ to the results from the dual simulation.
In the case of $\mu = 0.4$ (lhs.\ plot), the ITE produces reliable outcome for all values of $\tau$, while at $\mu = 1.0$ the ITE up to third order 
of $\rho$ starts to deviate from the dual variable data at  $\tau \sim 0.1812$.

\begin{figure}[t]
\begin{minipage}[b]{0.49\linewidth}
\centering
\includegraphics[height = 55mm]{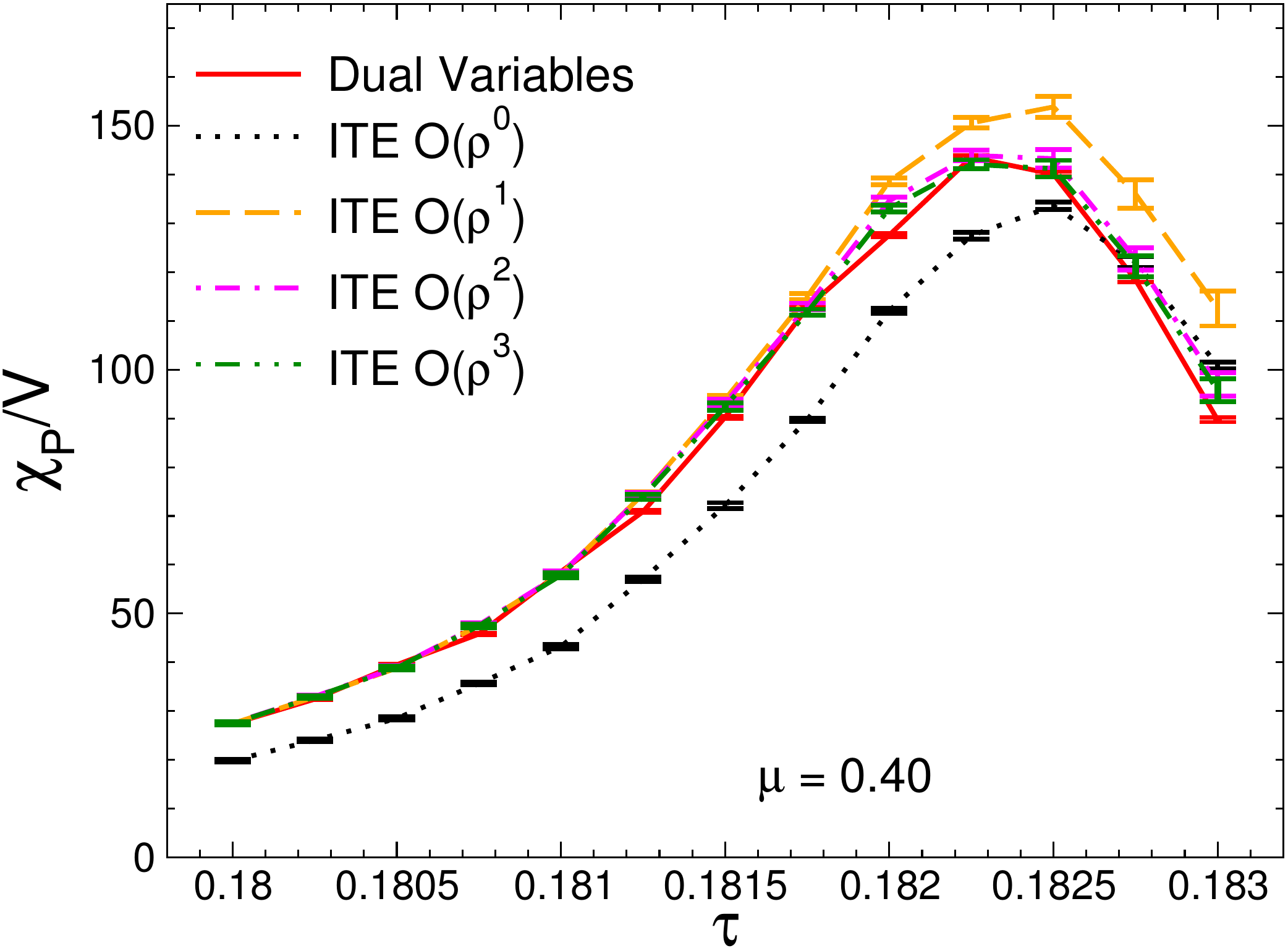}
\vspace*{-0.52cm}
\end{minipage}
\hspace{0.3cm}
\begin{minipage}[b]{0.47\linewidth}
\centering
\includegraphics[height = 55mm]{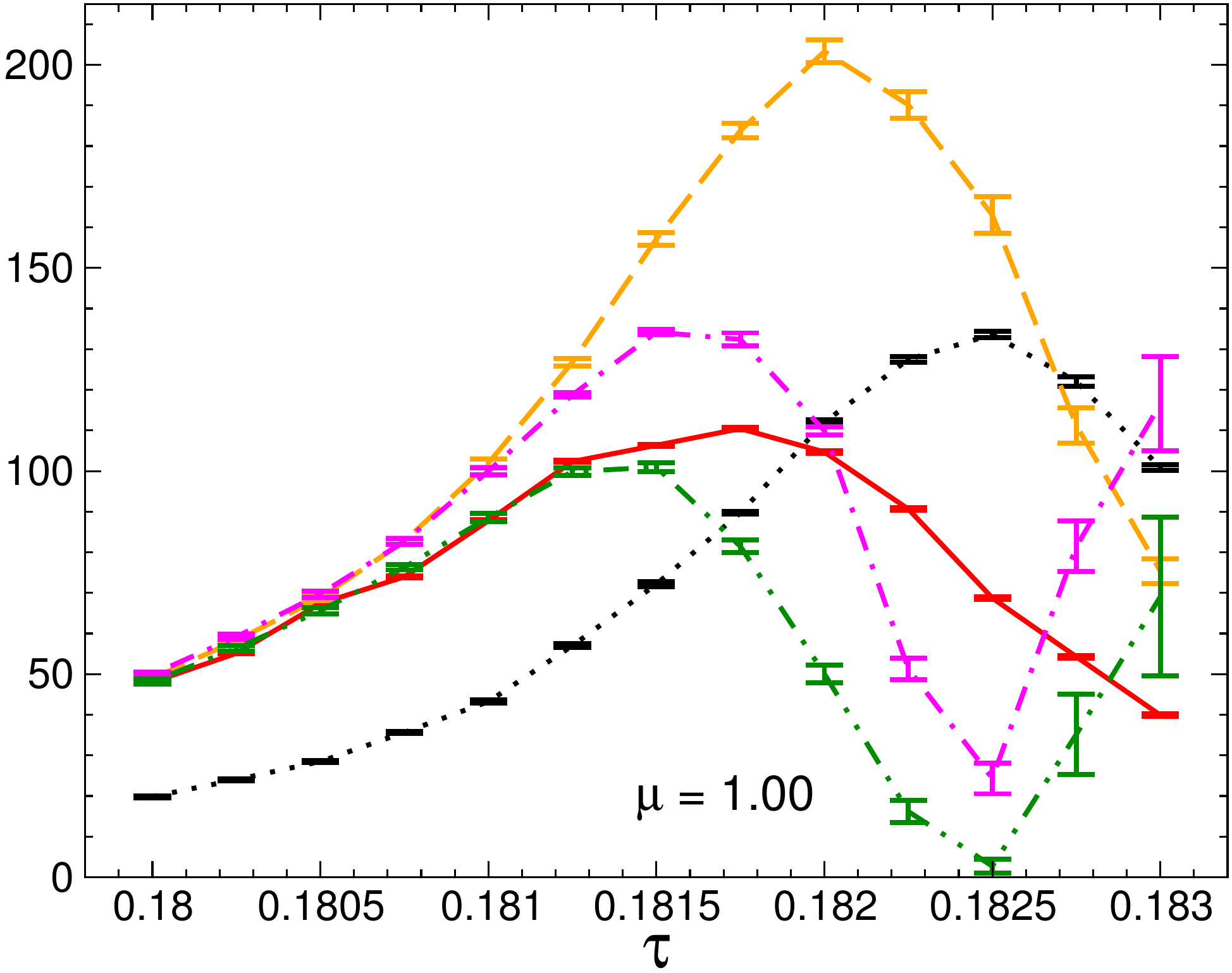}
\label{fig:chiP_k0001_ITE_dual}
\end{minipage}
\caption{Polyakov loop susceptibility obtained from the improved Taylor expansion for $\kappa = 0.001$ as a function of the temperature. 
Results up to third order of $\rho$ are shown and are compared to flux representation results on a $16^3$ lattice from $10^6$ measurements.
}
\label{fig:chiP_k0001_ITE_dual}
\end{figure}

\section{Direct comparison of all three expansion techniques}
\begin{figure}[t]
\begin{minipage}{0.48\textwidth}
\includegraphics[width=\textwidth]{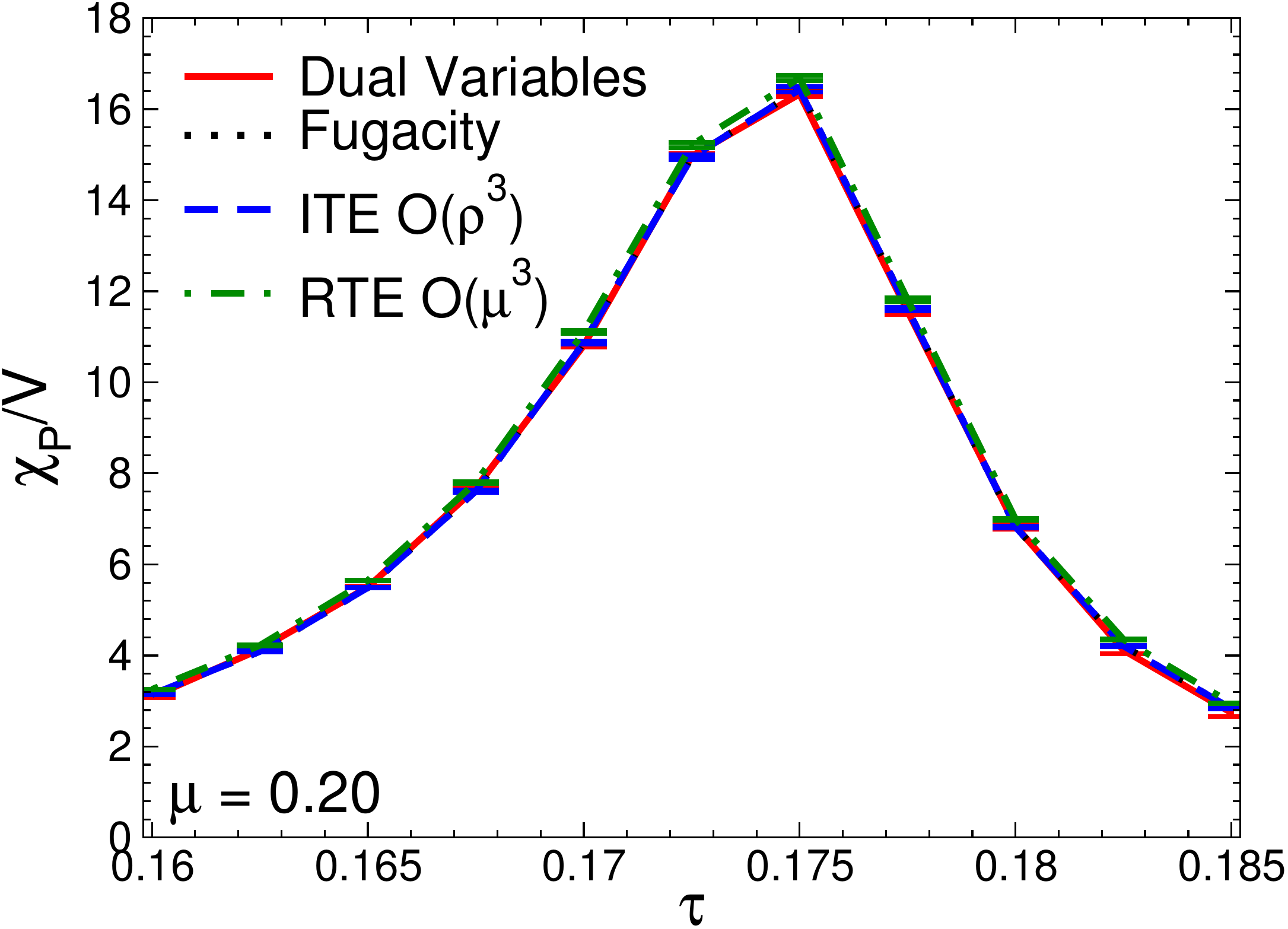} 
\end{minipage}
\begin{minipage}{0.48\textwidth}
\includegraphics[width=\textwidth]{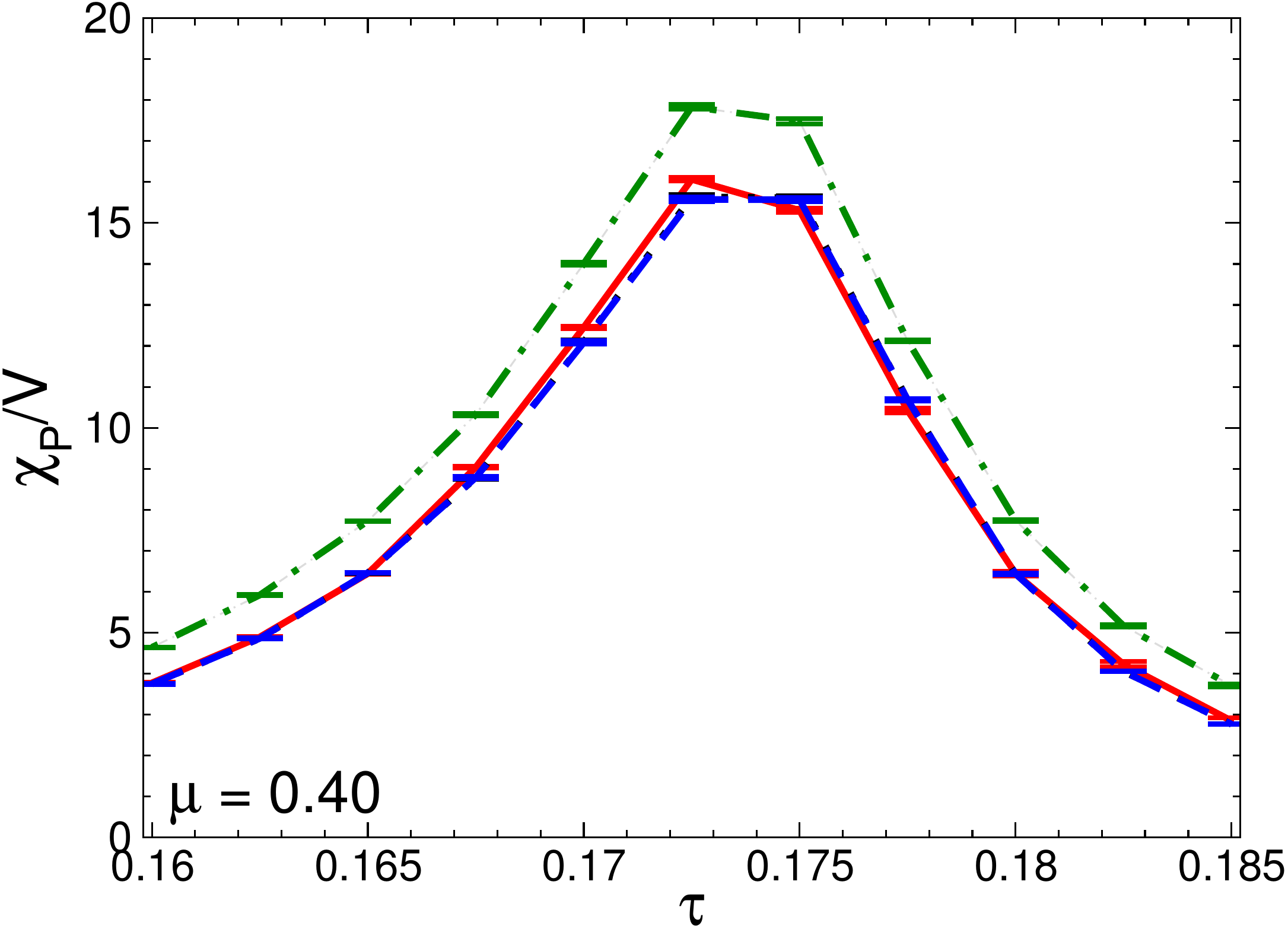}
\end{minipage}

\begin{minipage}{0.48\textwidth}
\includegraphics[width=\textwidth]{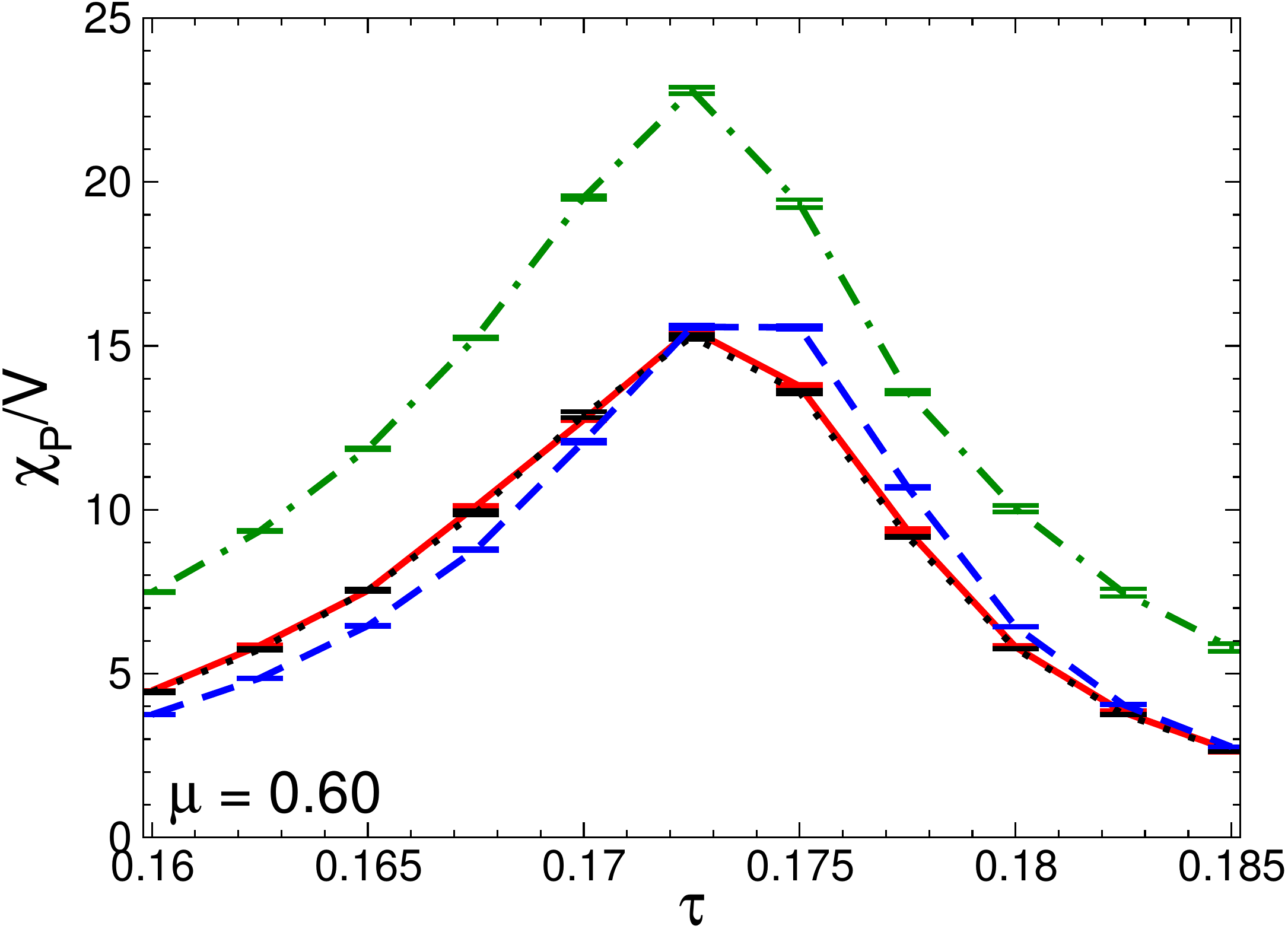}
\end{minipage}
\begin{minipage}{0.48\textwidth}
\includegraphics[width=\textwidth]{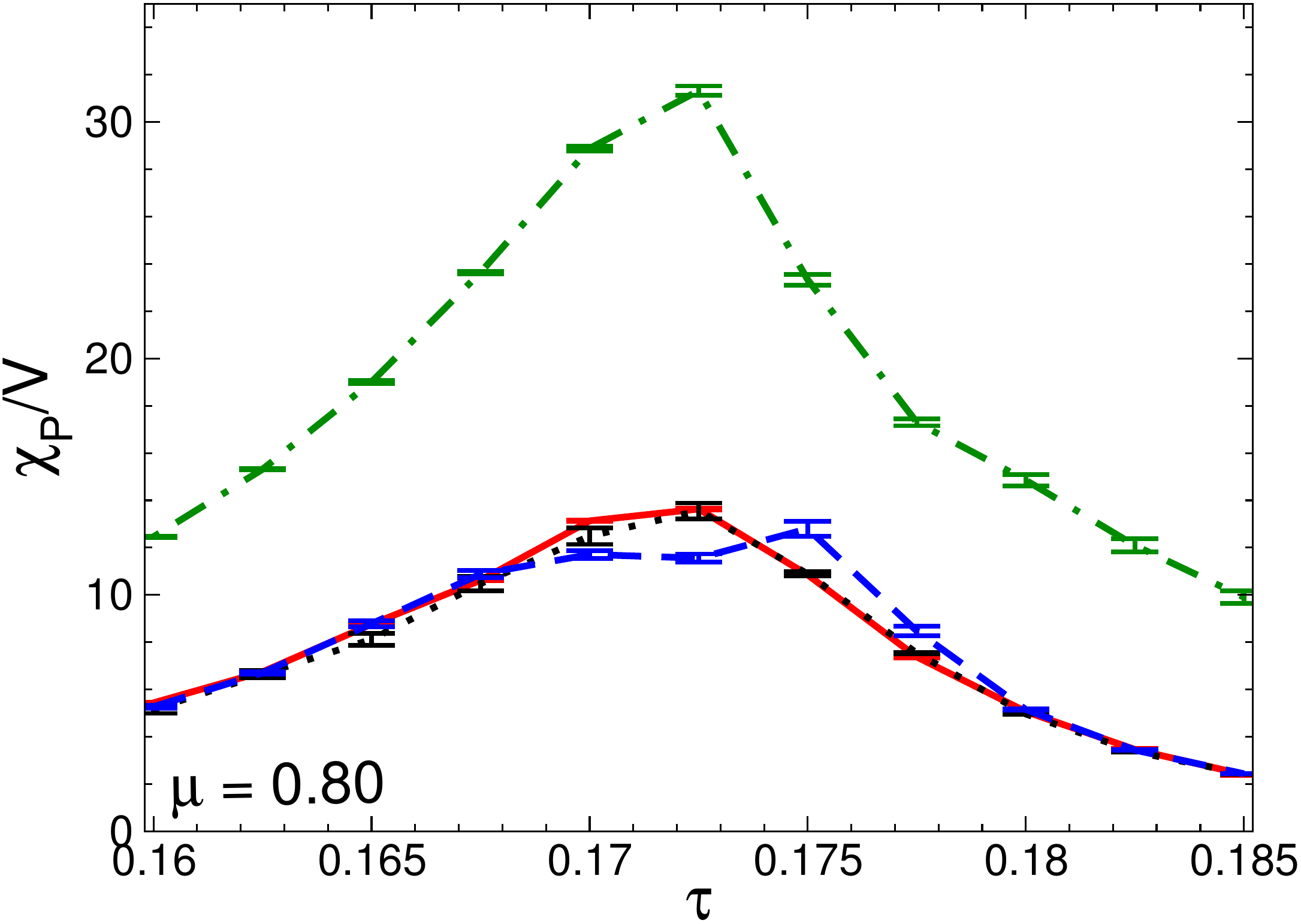}
\end{minipage}

\begin{minipage}{0.48\textwidth}
\includegraphics[width=\textwidth]{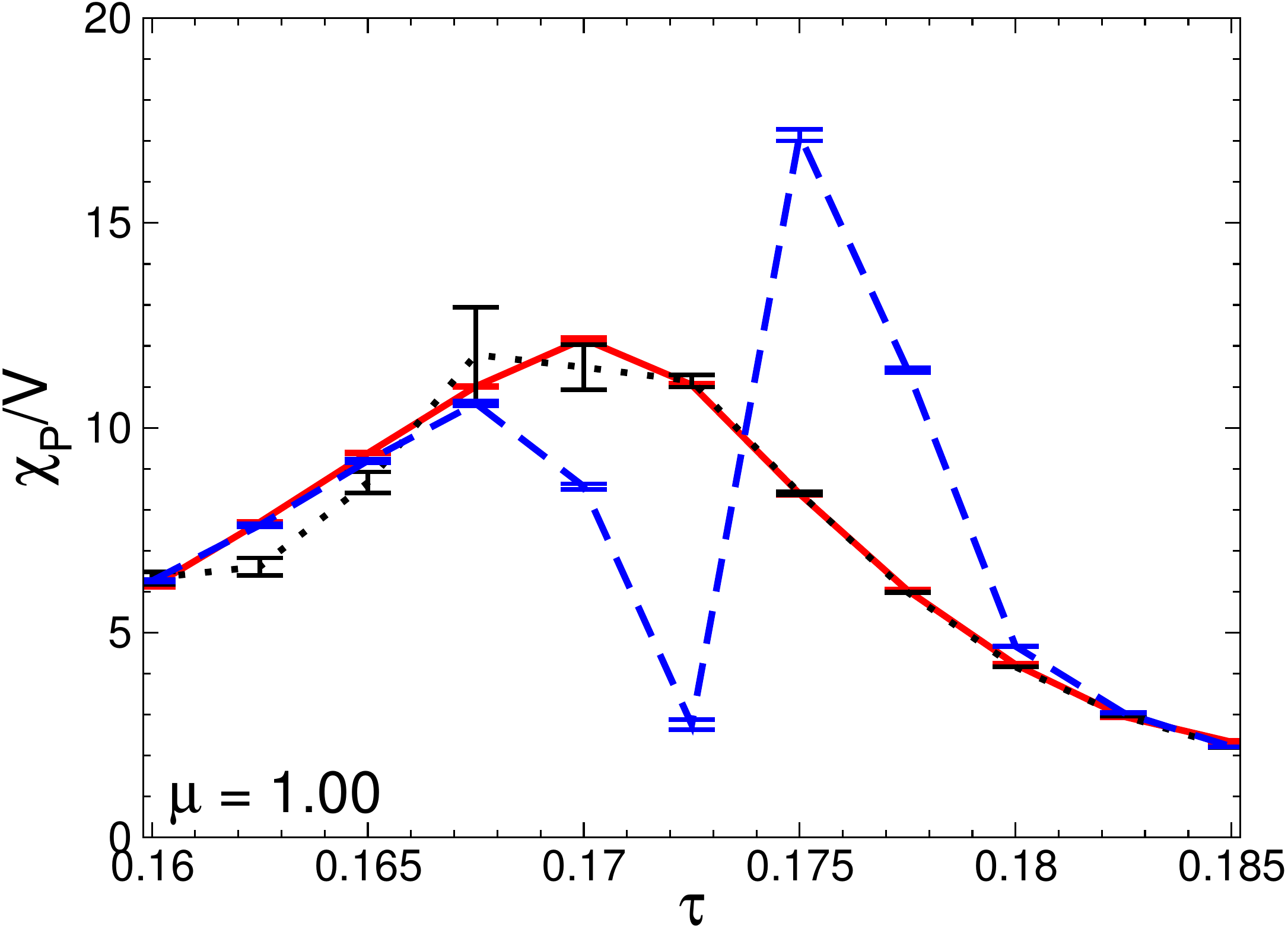}
\end{minipage}
\begin{minipage}{0.48\textwidth}
\includegraphics[width=\textwidth]{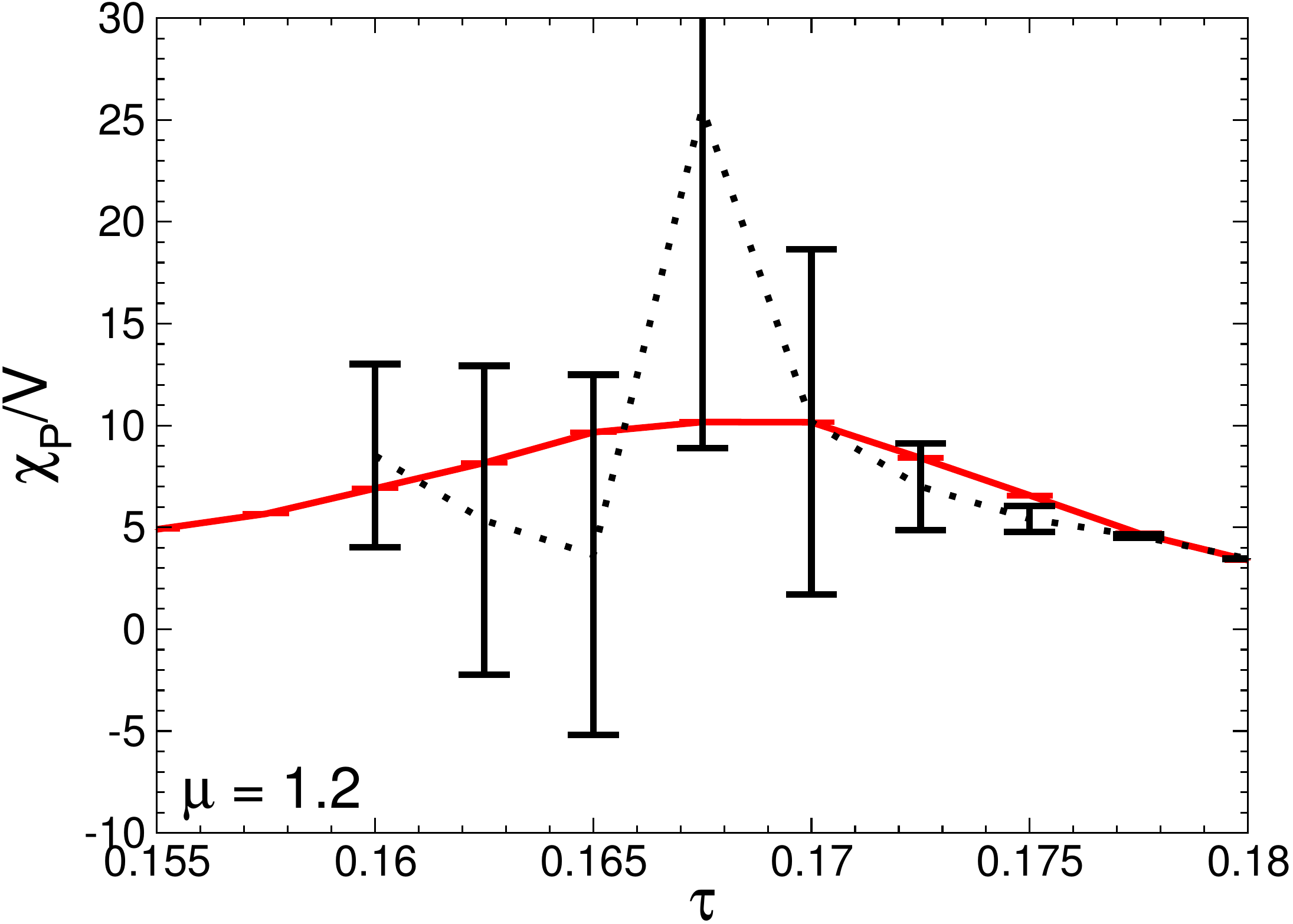}
\end{minipage}
\label{chiP_vergl}
\caption{Comparison of the Polyakov loop susceptibility at $ \kappa = 0.01$ and lattice size $16^3$ for fugacity, RTE and ITE 
expansions to results from a dual simulation for six values of the chemical potential $\mu$. }
\end{figure}

In Fig.~\ref{chiP_vergl} we systematically compare fugacity-, improved- and regular Taylor expansion (both up to 3-rd order) 
to results from flux representation for six 
values of $\mu$ at $\kappa = 0.01$. Dual variable data is from $10^6$ measurements, whereas fugacity and Taylor expansion need $4\cdot10^7$ 
configurations to produce reasonable data at $\mu \geq 1.0$. The RTE fails to converge already near $\mu = 0.6$, whereas the ITE produces reliable outcome 
up to $\mu = 0.8$. The fugacity expansion reproduces the results of the dual variables quite well, until the complex action problem becomes very severe 
at $\mu = 1.2$. 

\section{Summary}

In the work reported here, we compared fugacity-, regular Taylor- and improved Taylor-expansion concerning their reliability in reproducing the phase 
diagram of the effective center model \cite{centermodel1,centermodel2}. As reference data we used results from a simulation in the dual representation 
\cite{centermodel2} where the complex action problem is overcome for arbitrary values of $\mu$.
It turned out, that the loss of convergence of the fugacity expansion coincides with the range of chemical potential where the complex action problem 
becomes very severe (Fig.\ref{CAP}). For small values of $\kappa$, e.g. $\kappa = 0.001$, this leaves a relatively wide range of $\mu$-values that 
can be  explored reliably with the fugacity expansion. In comparison to that, the Taylor expansion methods are valid for only quite small values of the
chemical potential, where the ITE outperformed the RTE. None of the three expansion methods can reproduce the phase diagram of the 
effective center model to full extent. However, the assessment of their limitations in a comparison to the reliable reference data from the 
dual approach can be used to improve these series expansion techniques for QCD applications.

\end{document}